\documentclass[twocolumn,numberedappendix,trackchanges]{aastex62}

\usepackage{color}

\usepackage{mathtools}

\usepackage{bm}

\usepackage{float}

\usepackage{verbatim}

\newcommand{\msolar}{\mathrm{M}_\odot}

\newcommand{\amrex}{\texttt{AMReX}}

\newcommand{\castro}{\texttt{CASTRO}}

\newcommand{\microphysics}{\texttt{Microphysics}}

\newcommand{\matplotlib}{\texttt{matplotlib}}
\newcommand{\yt}{\texttt{yt}}

\begin{document}

\title{Numerical Stability of Detonations in White Dwarf Simulations}

\shorttitle{Detonation Stability}
\shortauthors{Katz and Zingale (2019)}

\author{Max P. Katz}
\affiliation
{
  NVIDIA Corporation, 2788 San Tomas Expressway, Santa Clara, CA, 95051, USA
}

\author{Michael Zingale}
\affiliation
{
  Department of Physics and Astronomy, Stony Brook University, Stony Brook, NY, 11794-3800, USA
}

\begin{abstract}
Some simulations of Type Ia supernovae feature self-consistent thermonuclear
detonations. However, these detonations are not meaningful if the simulations
are not resolved, so it is important to establish the requirements for achieving
a numerically converged detonation. In this study we examine a test detonation
problem inspired by collisions of white dwarfs. This test problem demonstrates
that achieving a converged thermonuclear ignition requires spatial resolution
much finer than 1 km in the burning region. Current computational resource
constraints place this stringent resolution requirement out of reach for
multi-dimensional supernova simulations. Consequently, contemporary simulations
that self-consistently demonstrate detonations are possibly not converged
and should be treated with caution.
\end{abstract}

\section{Introduction}
\label{sec:introduction}

Thermonuclear detonations are common to all current likely models of Type Ia
supernovae (SNe Ia), but how they are actually generated in progenitor systems
is still an open question. Different models predict different locations for
the detonation and different mechanisms for initiating the event. Common to all
of the cases is a severe lack of numerical resolution in the location where the
detonation is expected to occur. The length and time scale at which a detonation
forms is orders of magnitude smaller than the resolution that typical multi-dimensional
hydrodynamic simulations can achieve. The mere presence of a detonation (or lack thereof)
in a simulation is therefore only weak evidence regarding whether a detonation would truly occur.

In this study we examine the challenges associated with simulating thermonuclear detonations.
The inspiration for this work comes from the literature on head-on collisions of WDs,
which can occur, for example, in certain triple star systems \citep{thompson:2011,hamers:2013}.
WD collisions rapidly convert a significant amount of kinetic energy into thermal energy and
thus set up conditions ripe for a thermonuclear detonation. Since they are easy to set up in a simulation,
they are a useful vehicle for studying the properties of detonations.

Early studies on WD collisions \citep{rosswog:2009,raskin:2010,loren-aguilar:2010,
hawley:2012,garcia-senz:2013} typically had effective spatial resolutions in the burning region of
100--500 km for the grid codes, and 10--100 km for the SPH codes, and observed
detonations that convert a large amount of carbon/oxygen material into iron-group elements.
These studies varied in methodology (Lagrangian versus Eulerian evolution, nuclear network
used) and did not closely agree on the final result of the event (see Table 4 of
\cite{garcia-senz:2013} for a summary).

There is mixed evidence for simulation convergence presented in these studies.
\cite{raskin:2010} claim that their simulations are converged in nickel yield up to 2 million
(constant mass) particles, but the nickel yield still appears to be trending slightly upward
with particle count. The earlier simulations of \cite{raskin:2009} are not converged up to
800,000 particles, where the smoothing length was kept constant instead of the particle mass.
\cite{hawley:2012} do not achieve convergence over a factor of 2 in spatial resolution.
\cite{garcia-senz:2013} claim at least qualitative (though not strict absolute) convergence, but
their convergence test is only over a factor of 2 in particle count, which is a factor of $2^{1/3} = 1.3$
in spatial resolution (for constant mass particles). \cite{kushnir:2013} test convergence over an
order of magnitude in spatial resolution, and find results that appear to be reasonably well
converged for one of the two codes used (VULCAN2D), and results that are not converged for the
other code used (FLASH). \cite{papish:2015} claim convergence in nuclear burning up to 10\% at a
resolution of 5--10 km, but do not present specific data demonstrating this claim or precisely define
what is being measured. \cite{loren-aguilar:2010} and \cite{rosswog:2009} do not present convergence
studies for their work.

\cite{kushnir:2013} argued that many of these simulations featured numerically unstable evolution,
ultimately caused by the zone size being significantly larger than the length scale over which
detonations form. The detonation length scale can vary widely based on physical conditions
\citep{seitenzahl:2009,garg:2017} but is generally not larger than 10 km. \citeauthor{kushnir:2013}
argue that this numerically unstable evolution is the primary cause of convergence difficulties.
They further argue that it is possible to apply a burning limiter to achieve converged results,
which was used in their work and later the simulations of \cite{papish:2015}. We investigate
this hypothesis in \autoref{sec:unstable_burning}.

In this paper, we attempt to find what simulation length scale is required to achieve
converged thermonuclear ignitions. The inspiration for this work comes from our
simulations of WD collisions using the reactive hydrodynamics code \castro\
\citep{castro, astronum:2017}. We have done both 2D axisymmetric and 3D simulations
of collisions of $0.64\ \msolar$ carbon/oxygen WDs, and we were unable to achieved converged
simulations at any resolution we could afford to run (the best was an effective zone size of
0.25 km, using adaptive mesh refinement, for the 2D case). We were therefore forced to
turn to 1D simulations, where we can achieve much higher resolution (at the cost, of course,
of not being able to do a test that can be directly compared to multi-dimensional simulations).
We believe the simulations presented below help show why we and others had difficulty
achieving convergence at the resolutions achievable in multi-dimensional WD collision simulations.

\section{Test Problem}
\label{sec:collisions}

Our test problem is inspired by \cite{kushnir:2013}, and very loosely approximates the
conditions of two $0.64\ \msolar$ WDs colliding head-on. The simulation domain is 1D with a
reflecting boundary at $x = 0$. For $x > 0$ there is a uniform fluid composed (by mass)
of $50\%\, ^{12}$C, $45\%\, ^{16}$O, and $5\%\, ^{4}$He. The fluid is relatively cold,
$T = 10^7$ K, has density $\rho = 5 \times 10^6$ g/cm$^3$, and is traveling toward the
origin with velocity $-2 \times 10^8$ m/s. A uniform constant gravitational acceleration
is applied, $g = -1.1 \times 10^8$ m/s$^{2}$. This setup causes a sharp initial release
of energy at $x = 0$, and the primary question is whether a detonation occurs promptly
near this contact point, or occurs later (possibly at a distance from the contact point).
The simulated domain has width $1.6384 \times 10^9$ cm, and we apply inflow boundary conditions
that keep feeding the domain with material that has the same conditions as the initial fluid.
Simulations are performed with the adaptive mesh refinement (AMR) code \castro.
For the burning we use the alpha-chain nuclear network \texttt{aprox13}.
Release 18.12 of the \castro\ code was used. The \amrex\ and \microphysics\ repositories
that \castro\ depends on were also on release 18.12. The problem is located in the
\texttt{Exec/science/Detonation} directory, and we used the \texttt{inputs-collision} setup.

The simulation is terminated when the peak temperature on the domain first reaches
$4 \times 10^9$ K, which we call a thermonuclear ignition (for reference, the
density at the location where the ignition occurs is approximately $1.4\times 10^7\ \text{g / cm}^3$). This stopping criterion is a
proxy for the beginning of a detonation. Reaching this temperature does not guarantee
that a detonation will begin, and in this study we do not directly address the question
of whether a ignition of this kind always leads to a detonation. Nor are we commenting
on the physics of the ignition process itself. Rather, the main question
we investigate here is whether this ignition is numerically converged, and for this purpose
this arbitrary stopping point is sufficient, since in a converged simulation the stopping point
should be reached at the same time independent of resolution. A converged ignition
is a prerequisite to having a converged detonation. We measure two diagnostic quantities:
the time since the beginning of the simulation required to reach this ignition criterion,
and the distance from the contact point of the peak temperature.

The only parameter we vary in this study is the spatial resolution used for this problem.
For low resolutions we vary only the base resolution of the grid, up to a resolution of
0.25 km. For resolutions finer than this, we fix the base grid at a resolution of 0.25 km,
and use AMR applied on gradients of the temperature. We tag zones for refinement if the temperature
varies by more than 50\% between two zones. Timesteps are limited only by the hydrodynamic
stability constraint, with CFL number 0.5. Although this leads to Strang splitting error
in the coupling of the burning and hydrodynamics for low resolution, we have verified that
the incorrect results seen at low resolution do not meaningfully depend on the timestep constraint
(both by applying a timestep limiter based on nuclear burning, and by using the spectral deferred
corrections driver in \castro, which directly couples the burning and hydrodynamics). At very high
resolution, the splitting error tends to zero as the CFL criterion decreases the timestep.

\begin{figure}[ht]
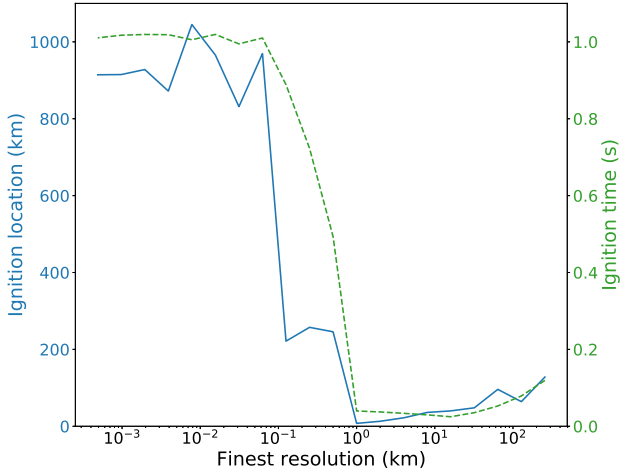

  \centering
  \includegraphics[scale=0.30]{{{amr_ignition_self-heat}}}
  \caption{Distance from the contact point of the ignition (solid blue), and time of the
           ignition (dashed green), as a function of finest spatial resolution.
           \label{fig:self-heat-distance}}
\end{figure}

\autoref{fig:self-heat-distance} shows our main results. The lowest resolution we consider,
256 km, is typical of the early simulations of white dwarf collisions, and demonstrates a
prompt ignition near the contact point. As the (uniform) resolution increases, the ignition
tends to occur earlier and nearer to the contact point. This trend is not physically meaningful:
all simulations with resolution worse than about 1 km represent the same prompt central ignition,
and as the resolution increases, there are grid points physically closer to the center that can ignite.
However, when the resolution is better than 1 km, the situation changes dramatically: the prompt
central ignition does not occur, but rather the ignition is delayed and occurs further from the contact
point. When we have finally reached the point where the curves start to flatten and perhaps begin to converge, 
the ignition occurs around 900 km from the contact point, about 1 second after contact (contrast to less than
0.05 seconds for the simulation with 1 km resolution). Even at this resolution, it is not clear if
the simulation is converged. We were unable to perform higher resolution simulations to check convergence
due to the length of time that would be required.

\begin{figure}[ht]
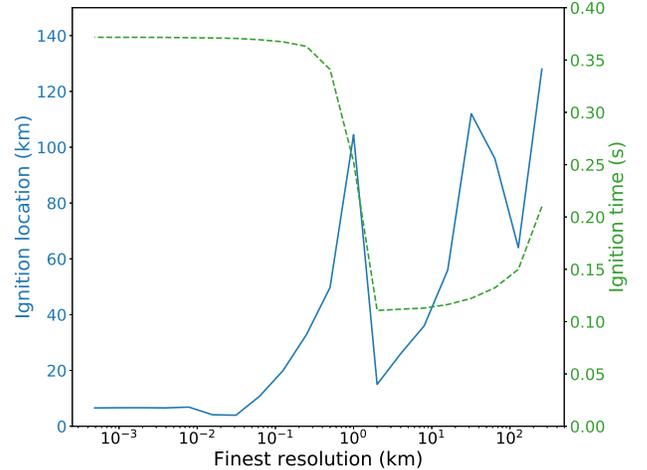

  \centering
  \includegraphics[scale=0.30]{{{amr_ignition_co}}}
  \caption{Similar to \autoref{fig:self-heat-distance}, but with pure C/O material.
           Note the different vertical axis scale.
           \label{fig:self-heat-distance-co}}
\end{figure}

We also tested a similar configuration made of pure carbon/oxygen material (equal fraction by mass).
This is closer to the configuration used in the $0.64\ \msolar$ WD collision simulations that previous
papers have focused on. However, for the setup described above, pure carbon/oxygen conditions do not
detonate at all. This is not particularly surprising, since the 1D setup is a very imperfect representation
of the real multi-dimensional case, and is missing multi-dimensional hydrodynamics that could substantially
alter the dynamical evolution. So the small amount of helium we added above ensured that the setup ignited.
(Of course, there will likely be a small amount of helium present in C/O white dwarfs as a remnant of the prior
stellar evolution.) However, we can prompt the C/O setup to ignite by starting the initial temperature
at $10^9$ K instead of $10^7$ K. This loosely mimics the effect from the first test where helium burning
drives the temperature to the conditions necessary to begin substantial burning in C/O material. But since
no helium is present in this case, it allows us to test whether it is easier to obtain convergence for pure
C/O burning, even though the test itself is artificial. The only other change relative to the prior test is
that we refined on relative temperature gradients of 25\% instead of 50\%. The results for this case are shown in
\autoref{fig:self-heat-distance-co}. In this case, the ignition is central at all resolutions, but the
simulation is still clearly unconverged at resolutions worse than 100 m, as the ignition becomes significantly
delayed at high resolution.

This story contains two important lessons. First, the required resolution for even a qualitatively converged
simulation, less than 100 m, is out of reach for an analogous simulation done in 3D. Second, the behavior for 
resolutions worse than 1 km qualitatively appears to be converged, and one could perhaps be misled into
thinking that there was no reason to try higher resolutions, which is reason for caution in interpreting
reacting hydrodynamics simulations. With that being said, our 1D tests are not directly comparable to
previous multi-dimensional WD collision simulations. The 1D tests should not be substituted for understanding the
actual convergence properties of the 2D/3D simulations, which may have different resolution requirements for
convergence. Our tests suggest only that it is plausible that simulations at kilometer-scale (or worse) resolution
are unconverged. This observation is, though, consistent with the situation described in \autoref{sec:introduction},
where our 2D WD collision simulations (not shown here) are unconverged, and many of the previous collision simulations
presented in the literature have relatively weak evidence for convergence.

\section{Numerically Unstable Burning}
\label{sec:unstable_burning}

\citet{kushnir:2013} observe an important possible failure mode
for reacting hydrodynamics simulations. Let us define $\tau_{\rm e} = e / \dot{e}$
as the nuclear energy injection timescale, and $\tau_{\rm s} = \Delta x / c_{\rm s}$
as the sound-crossing time in a zone (where $\Delta x$ is the grid
resolution and $c_{\rm s}$ is the speed of sound). When the sound-crossing
time is too long, energy is built up in a zone faster than it can be
advected away by pressure waves. This effect generalizes to
Langrangian simulations as well, where $\tau_{\rm s}$ should be understood
as the timescale for transport of energy to a neighboring fluid element.
This is of course a problem inherent 
only to numerically discretized systems as the underlying fluid equations
are continuous. This can lead to a numerically seeded detonation
caused by the temperature building up too quickly in the zone. The
detonation may be spurious in this case. If $\tau_{\rm s} \ll \tau_{\rm e}$,
we can be confident that a numerically seeded detonation has not
occurred. In practice, we quantify this requirement as:
\begin{equation}
  \tau_{\rm s} \leq f_{\rm s}\, \tau_{\rm e} \label{eq:burning_limiter}
\end{equation}
and require that $f_{s}$ is sufficiently smaller than one.
\citet{kushnir:2013} state that $f_{\rm s} = 0.1$ is a sufficient
criterion for avoiding premature ignitions. \citeauthor{kushnir:2013}
enforced this criterion on their simulations by artificially limiting
the magnitude of the energy release after a burn, and claimed that
this is resulted in more accurate WD collision simulations.

\begin{figure}[ht]
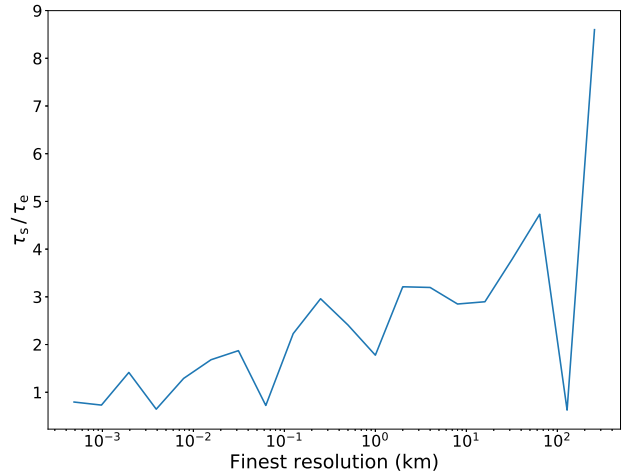

  \centering
  \includegraphics[scale=0.30]{{{amr_ignition_self-heat_ts_te}}}
  \caption{Ratio of the sound-crossing timescale to the energy injection timescale
           for the simulations in \autoref{fig:self-heat-distance}.
           \label{fig:self-heat-ts_te}}
\end{figure}

We find that for our test problem (and also the WD collisions we have simulated)
we do observe $\tau_{\rm s} > \tau_{\rm e}$; typically the ratio is a factor of 2--5 at
low resolution (see \autoref{fig:self-heat-ts_te}). This means that an ignition is
very likely to occur for numerical reasons, regardless of whether it would occur for physical reasons.
At low resolution, adding more resolution does not meaningfully improve the ratio of
$\tau_s$ to $\tau_e$ at the point of ignition. The ignition timescale is so short
that almost all of the energy release occurs in a single timestep even though the
timestep gets shorter due to the CFL limiter. It is only when the resolution gets
sufficiently high that we can simultaneously resolve the energy release over multiple
timesteps and the advection of energy across multiple zones. Even at the highest resolution
we could achieve for the test including helium, about 50 cm, $\tau_s / \tau_e$ was 0.8 at ignition, which is not
sufficiently small to be confident of numerical stability. Note
that merely decreasing the timestep (at fixed resolution) does not help here either, as the
instability criterion is, to first order, independent of the size of the timestep.

We thus investigate whether limiting the energy release of the burn (we will term this
``suppressing'' the burn), as proposed by \citeauthor{kushnir:2013},
is a useful technique for avoiding the prompt detonation. Since the limiter ensures
the inequality in \autoref{eq:burning_limiter} holds by construction, the specific
question to ask is whether the limiter achieves the correct answer and is converged
in cases where the simulation would otherwise be uncorrect or unconverged.

Before we examine the results, consider a flaw in the application of the limiter:
a physical detonation may \textit{also} occur with the property that, in the detonating
zone, $\tau_s > \tau_e$. For example, consider a region of WD material at uniformly
high temperature, say $5 \times 10^9\ \text{K}$, with an arbitrarily large size,
say a cube with side length 100 km. This region will very likely ignite,
even if it is surrounded by much cooler material. By the time the material on
the edges can advect heat away, the material in the center will have long since
started burning carbon, as the sound crossing time scale is sufficiently large
compared to the energy injection time scale. This is true regardless of whether
the size of this cube corresponds to the spatial resolution in a simulation.
Suppression of the burn in this case is unphysical: if we have a zone matching
these characteristics, the zone should ignite.

When the resolution is low enough, there is a floor on the size of a hotspot,
possibly making such a detonation more likely. This is an unavoidable consequence
of the low resolution; yet, it may be the correct result of the simulation that
was performed. That is, even if large hotspots are unphysical because in reality
the temperature distribution would be smoother, if such a large hotspot \textit{were}
to develop (which is the implicit assumption of a low resolution simulation), then
it would likely ignite. If the results do not match what occurs at higher
resolution, then the simulation is not converged and the results are not reliable.
However, it may also be the case that a higher resolution simulation will yield
similar results, for example because even at the higher resolution, the physical
size of the hotspot stays the same. For this reason, an appeal to the numerical
instability criterion alone is insufficient to understand whether a given ignition
is real.

\begin{figure}[ht]
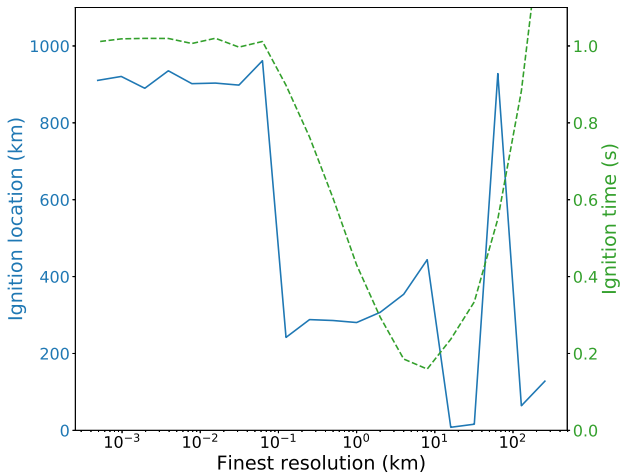

  \centering
  \includegraphics[scale=0.30]{{{amr_ignition_suppressed}}}
  \caption{Similar to \autoref{fig:self-heat-distance}, but for simulations with the
           suppressed burning limiter applied (\autoref{eq:burning_limiter}).
           \label{fig:suppressed-distance}}
\end{figure}

\autoref{fig:suppressed-distance} shows the results we obtain for our implementation of
a ``suppressed'' burning mode. In a suppressed burn, we limit the changes to the state
so that \autoref{eq:burning_limiter} is always satisfied. This is done by rescaling the
energy release and species changes from a burn by a common factor such that the equality
in \autoref{eq:burning_limiter} is satisfied. (If the inequality is already satisfied,
then the integration vector is not modified.) We find that the suppressed burn
generally does not yield correct results for low resolutions. The 64 km resolution
simulation happens to yield approximately the correct ignition distance, but it does
not occur at the right time, and in any case the incorrectness of the results at neighboring
resolutions suggests that this is not a robust finding. The suppressed burning simulation
reaches qualitative convergence at around the same 100 m resolution as the normal self-heating
burn. Because of both the theoretical reasons discussed above, and this empirical finding that
the burning suppression does not make low resolution simulations any more accurate, we do not
believe that the suppressed burning limiter should be applied in production simulations.
\newline 

\section{Conclusion}\label{Sec:Conclusion}
\label{sec:conclusion}

Our example detonation problem demonstrates, at least for this class of
hydrodynamical burning problem, a grid resolution requirement much more stringent
than 1 km. This test does not, of course, represent all possible WD burning conditions.
However, the fact that it is even possible for burning in white dwarf material to require a
resolution better than 100 m should suggest that stronger demonstrations of convergence are
required. This is especially true bearing in mind our observation that the numerical
instability can result in simulations that appear qualitatively converged when the
resolution is increased by a factor of one or two orders of magnitude but not three
orders of magnitude.

This study does not directly address the problem of how, in the detailed
microphysical sense, a detonation wave actually begins to propagate, as
we cannot resolve this length scale even in our highest resolution simulations.
Rather, we are making the point that for simulations in which a macroscopic
detonation wave appears self-consistently, this is only a valid numerical result
if the resolution is sufficiently high. This convergence requirement does
not imply that the detonation itself is physically realistic; but, it does
imply that we are not even correctly solving the fluid equations we intend
to solve when the convergence requirement is not met. We believe that our
test case can be useful in the future for testing algorithmic innovations
that hope to improve the realism of burning at low resolutions.

\acknowledgments

This research was supported by NSF award AST-1211563 and DOE/Office of
Nuclear Physics grant DE-FG02-87ER40317 to Stony Brook. An award of
computer time was provided by the Innovative and Novel Computational
Impact on Theory and Experiment (INCITE) program.  This research used
resources of the Oak Ridge Leadership Computing Facility located in
the Oak Ridge National Laboratory, which is supported by the Office of
Science of the Department of Energy under Contract
DE-AC05-00OR22725. Project AST106 supported use of the ORNL/Titan
resource.  This research used resources of the National Energy
Research Scientific Computing Center, which is supported by the Office
of Science of the U.S. Department of Energy under Contract
No. DE-AC02-05CH11231. The authors would like to thank Stony Brook
Research Computing and Cyberinfrastructure, and the Institute for
Advanced Computational Science at Stony Brook University for access
to the high-performance LIred and SeaWulf computing systems, the latter
of which was made possible by a \$1.4M National Science Foundation grant (\#1531492).

The authors thank Chris Malone and Don Willcox for useful discussions
on the nature of explosive burning, and Doron Kushnir for providing
clarification on the nature of the burning limiter used in \cite{kushnir:2013}.

This research has made use of NASA's Astrophysics Data System 
Bibliographic Services.

\facilities{OLCF, NERSC}
\software{\castro\ \citep{castro, astronum:2017},
          \amrex\ \citep{boxlib-tiling},
          \yt\ \citep{yt},
          \matplotlib\ \citep{matplotlib}
          }

\bibliographystyle{aasjournal}

\end{document}